\documentclass[twocolumn]{jpsj2w}
\def\runtitle{
H$_2$ Dissociative Adsorption at the Armchair Edges of Graphite
}

\title{%
H$_2$ Dissociative Adsorption at the Armchair Edges of Graphite
}

\author{%
Wilson Agerico {\sc Di\~no},$^{1,2,3}$ 
Yoshio {\sc Miura},$^{1,}$\thanks{Now at RIEC, Tohoku University, Sendai 980-8577, Japan.}
Hiroshi {\sc Nakanishi},$^{1}$ 
Hideaki {\sc Kasai}$^{1,}$\thanks{Corresponding author. E-mail: kasai@dyn.ap.eng.osaka-u.ac.jp.}
Tsuyoshi {\sc Sugimoto},$^{4}$ 
and Takuya {\sc Kondo}$^{4}$
}

\inst{%
$^{1}$Department of Applied Physics, Osaka University, Suita, Osaka 565-0871, Japan\\
$^{2}$Japan Science and Technology Corporation, Kawaguchi, Saitama 332-0012, Japan\\
$^{3}$Physics Department, De La Salle University, Taft Ave., 1004 Manila, Philippines\\
$^{4}$Toyota Motor Corporation, Toyota-cho, Aichi 471-8572, Japan\\
}

\recdate{10 November 2003}

\abst{%
We investigate and discuss how hydrogen behaves at the edges of a graphite sheet,  in particular the armchair edge.  Our density functional theory-based calculations results show that, in contrast to the zigzag edge [cf., e-J.\ Surf.\ Sci.\ Nanotech.\ {\bf 2} (2004) 77], regardless of orientation, there is an activation barrier hindering H$_2$ dissociation at the armchair edges.  And once they do get dissociatively adsorbed at the armchair edges, we find that it would be extremely hard to desorb the H from their adsorption sites at the armchair edges.  Furthermore, we also found that, consistent with our earlier conclusions [cf., J.\ Phys.\ Soc.\ Jpn.\ {\bf 72} (2003) 1867], it is unlikely that we would find a whole H$_2$ in between plain graphite sheets.
}

\kword{%
hydrogen, deuterium, atomic hydrogen, molecular hydrogen, carbon nanomaterials, nanographite, graphite layers, armchair edge, zigzag edge, density functional theory, neutron diffraction experiments, thermal desorption experiments
}

\begin{document}
\sloppy
\maketitle

Carbon-based nanomaterials, e.g., carbon nanotubes (CNTs) and graphite nanofibers (GNFs), have been attracting much attention because of their purported potentials as materials for gas storage.  Earlier reports of high hydrogen uptake of these materials make them attractive as hydrogen storage devices in fuel-cell-powered electric vehicles~\cite{Cheng-Carbon2001,Froudakis-JPCM2002}.  However, although various theoretical and experimental studies have been made  (cf.,~\citen{Chen-SS1989,Dillon-Nature1997,Wang-JCP1999,Darkrim-JCP1998,Chambers-JPCB1998,Jeloaica-CPL1999,Simonyan-JCP1999,Liu-Science1999,Lee-APL2000,Williams-CPL2000,Tada-PRB2001,Orimo-JAP2001,Rutigliano-CPL2001,Yang-Carbon2002,Fukunaga-JAC2001,Sha-SS2002}, and references therein), and initial reports seem to show promising and spectacular results (more than 60 wt\% in some cases, which is way above the minimum 6.5 wt\% hydrogen storage density requirement for economically feasible mobile application of hydrogen fuel at room temperature), research in this field remains contentious, as later studies reveal trends significantly less encouraging than were first contemplated~\cite{schlapbach-zuttel,klitzing}.  This situation is indicative of  the necessity for more systematic investigations.  On top of that, there is a necessity to develop a microscopic picture of the mechanism underlying hydrogen adsorption onto, absorption into, and desorption from carbon related materials.  With this in mind, we have done, and are still continuing, a series of studies on the interaction of hydrogen with various surfaces of graphite~\cite{h2-c,h-cnt,h-c,h-cc,h-ze}.

Invoking the density functional theory (DFT), we calculated the potential energy surfaces (PES) relevant to the dissociative adsorption of H$_2$ onto a graphite sheet~\cite{h2-c}.  Our results show that the reconstruction of the C atoms plays an important role in understanding the H$_2$-graphite surface interaction.  We observed a lowering of the dissociation barrier by ca.\ 1 eV, as a result of the relaxation outwards of the C atoms (changing its structure from $sp^2$ to $sp^3$-like), to meet the incoming H$_2$.  However, this is not enough to allow H$_2$ to dissociate, as the dissociation barriers still remains high at ca.\ 3~eV.  Furthermore, because the C atoms have to move out and meet the incoming H (H$_2$), this becomes difficult for H$_2$ dissociation, esp., when the C atoms involved are very close to each other (ca.\ 1.42~\AA).  The PES for H$_2$ dissociation on the surface plane of a graphite sheet is thus azimuthally corrugated.  The relaxation of the C atoms also stabilizes the H-graphite surface interaction, with a corresponding binding energy of 0.67~eV.  We were later informed~\cite{zecho-pc} that the theoretical results quantitatively agree with recent experimental results~\cite{zecho-jcp}.  

These results also give us an idea as to how hydrogen would behave when it interacts with a single-walled  carbon nanotube (SWCNT)~\cite{h-cnt}.  As expected from earlier results~\cite{h2-c,h-c}, a SWCNT with a small diameter adsorbs H quite easily.  However, as the diameter increases, the potential energy features approach that for hydrogen interaction with the surface plane of a graphite sheet.  It becomes more and more difficult to dissociate H$_2$ as the SWCNT diameter increases.

We also considered the possibility of finding hydrogen in between graphite layers~\cite{h-cc}.  Again, the reconstruction of the C atoms played an important role in determining what stable configurations hydrogen will assume once found inside/between graphite layers---the H atom assumes a position ca.\ 1~\AA~from one C atom on one sheet and ca.\ 2~\AA~from the C atom on the other sheet.  On the other hand, it would be difficult to dissociate H$_2$ in between graphite layers, let alone find them as H$_2$ between the graphite layers.  Again, we found amazing agreement between our results and recent observations from neutron diffraction~\cite{Fukunaga-JAC2001} and thermal desorption~\cite{Orimo-JAP2001} studies on deuterated nanographite.

Having found that it would be difficult for the H atom to reach inside/between the graphite layers through the hexagonal holes at the surface planes~\cite{h2-c,h-c}, and that it would also be difficult to find H$_2$ inside/between graphite layers~\cite{h-cc}, we considered the next probable surfaces on graphite, viz., the edges.   Our calculation results show that the zigzag edge is very reactive~\cite{h-ze}, and it is possible to dissociatively adsorb H$_2$  at the zigzag edge without any activation barrier hindering the reaction. We continue our study here, and consider the interaction of hydrogen with the other remaining edge of graphite, viz., the armchair edge.

\begin{figure}[tbp]
\begin{center}
\includegraphics[width=\linewidth]{Fig-1.epsf}
\end{center}
\caption{({\bf Upper Panel}) H$_2$ configuration with respect to the armchair edge, the H-H bond is oriented parallel to and immediately above one of the graphite sheets as indicated.  H$_2$ is constrained to dissociate over the indicated site in a parallel orientation and a planar geometry.  ({\bf Lower Left Panel}) Contour plot of the calculated potential energy surface (PES) for H$_2$ dissociation at the indicated site on the armchair edge of graphite, as a function of the H-H separation $r$, and the normal distance $Z$ of the H$_2$ center-of-mass from the armchair edge. The dashed-line traces the {\it path of least resistance/potential } (reaction path).  The interlayer distance between graphite sheets is ca.\ 3.5~\AA.  ({\bf Lower Right Panel}) Calculated potential energy along the reaction path, i.e., along dashed-line indicated on the PES shown on the {\bf Lower Left Panel}.  Energies are given in electronvolts relative to the values at ($Z=\infty$, $r\approx 0.74$~\AA).  Contour spacing is $\approx 0.2$~eV.}
\label{fig1}
\end{figure}

\begin{figure}[tbp]
\begin{center}
\includegraphics[width=\linewidth]{Fig-2.epsf}
\end{center}
\caption{({\bf Upper Panel}) H$_2$ configuration with respect to the armchair edge, the H-H bond is oriented parallel to and immediately above one of the graphite sheets as indicated.  H$_2$ is constrained to dissociate over the indicated site in a parallel orientation and a planar geometry.  ({\bf Lower Left Panel}) Contour plot of the calculated potential energy surface (PES) for H$_2$ dissociation at the indicated site on the armchair edge of graphite, as a function of the H-H separation $r$, and the normal distance $Z$ of the H$_2$ center-of-mass from the armchair edge. The dashed-line traces the {\it path of least resistance/potential } (reaction path).  The interlayer distance between graphite sheets is ca.\ 3.5~\AA.  ({\bf Lower Right Panel}) Calculated potential energy along the reaction path, i.e., along dashed-line indicated on the PES shown on the {\bf Lower Left Panel}.  Energies are given in electronvolts relative to the values at ($Z=\infty$, $r\approx 0.74$~\AA).  Contour spacing is $\approx 0.2$~eV.}
\label{fig2}
\end{figure}

We perform DFT-based total energy calculations, using the plane waves and pseudopotentials~\cite{dacapo}. The graphite supercell used in the calculations consists of two graphite sheets/layers, which are separated by ca.\ 3.5~\AA~and shifted relative to each other as shown on the Upper Panels of Figs.~\ref{fig1} to \ref{fig5}.   Each graphite layer consists of 21 carbon atoms with a nearest neighbor distance of $1.42$~\AA~. We have chosen a sufficiently large (graphite) supercell to avoid interactions between the H atoms (H$_{2}$ molecules) in the neighboring supercells. The electron-ion interaction is described by optimized ultrasoft pseudopotentials~\cite{Vanderbilt-PRB1990}, and the Kohn-Sham equations are solved using plane waves with kinetic energies up to $\approx 476$~eV.  The surface Brillouin zone integration is performed using the special-point sampling technique of Monkhorst and Pack (with $4 \times 4 \times 1$ sampling meshes)~\cite{Monkhorst-PRB1976}.   For the exchange correlation energy, we adopt the generalized gradient approximation (GGA)~\cite{Rasolt-PRB1986,Perdew-PRB1992}.  No significant change in the numerical results was observed when we increased the kinetic energy cutoff and the number of sampling points.  

Figures~\ref{fig1} and \ref{fig2} (Lower Left Panels) show the calculated potential energy contours for H$_2$ dissociation on the armchair edge of graphite, as a function of the H-H separation $r$, and the normal distance $Z$ of the H$_2$ center-of-mass from the armchair edge.  The H$_2$ is initially oriented with the H-H bond parallel to one of the graphite sheets, and the dissociation is constrained in a parallel orientation with respect to the armchair edge of one of the graphite sheets as indicated in the Upper Panels of Figs.~\ref{fig1} and \ref{fig2}.  H$_2$ dissociation is constrained in a parallel orientation with respect to the armchair edge, and a planar geometry.  Energies are given in electronvolts relative to the values for the case when H$_2$ is in the gas phase, far ($Z=\infty$) from the graphite. The interlayer distance between graphite sheets is ca.\ 3.5~\AA.   

We can see that there is an activation barrier hindering H$_2$ dissociation on the armchair edge of graphite.  There is ca.\ a 1.7 eV barrier hindering H$_2$ dissociation in the configuration depicted in Fig.~\ref{fig1} (cf., corresponding Lower Right Panel), and ca.\ 0.6 eV hindering H$_2$ dissociation for the configuration depicted in Fig.~\ref{fig2} (cf., corresponding Lower Left Panel).  Furthermore, assuming that we do succeed in dissociating H$_2$ at the indicated sites, we encounter the problem of how to desorb these H atoms, as they stick strongly to the corresponding edges, with an adsorption energy ranging from ca.\ 4 to 5 eV.

\begin{figure}[tbp]
\begin{center}
\includegraphics[width=\linewidth]{Fig-3.epsf}
\end{center}
\caption{({\bf Upper Right Panel}) H$_2$ configuration with respect to the armchair edge, the H-H bond is oriented perpendicular to and in between the two graphite sheets as indicated.  H$_2$ is constrained to dissociate over the indicated site in a parallel orientation and a planar geometry.  ({\bf Left Panel}) Contour plot of the calculated potential energy surface (PES) for H$_2$ dissociation at the indicated site on the armchair edge of graphite, as a function of the H-H separation $r$, and the normal distance $Z$ of the H$_2$ center-of-mass from the armchair edge. The dashed-line traces the {\it path of least resistance/potential } (reaction path).  The interlayer distance between graphite sheets is ca.\ 3.5~\AA.  ({\bf Lower Right Panel}) Calculated potential energy along the reaction path, i.e., along dashed-line indicated on the PES shown on the {\bf Left Panel}.  Energies are given in electronvolts relative to the values at ($Z=\infty$, $r\approx 0.74$~\AA).  Contour spacing is $\approx 0.2$~eV.}
\label{fig3}
\end{figure}

\begin{figure}[t]
\begin{center}
\includegraphics[width=\linewidth]{Fig-4.epsf}
\end{center}
\caption{({\bf Upper Right Panel}) H$_2$ configuration with respect to the armchair edge, the H-H bond is oriented perpendicular to and in between the two graphite sheets as indicated.  H$_2$ is constrained to dissociate over the indicated site in a parallel orientation and a planar geometry.  ({\bf Left Panel}) Contour plot of the calculated potential energy surface (PES) for H$_2$ dissociation at the indicated site on the armchair edge of graphite, as a function of the H-H separation $r$, and the normal distance $Z$ of the H$_2$ center-of-mass from the armchair edge. The dashed-line traces the {\it path of least resistance/potential } (reaction path).  The interlayer distance between graphite sheets is ca.\ 3.5~\AA.  ({\bf Lower Right Panel}) Calculated potential energy along the reaction path, i.e., along dashed-line indicated on the PES shown on the {\bf Left Panel}.  Energies are given in electronvolts relative to the values at ($Z=\infty$, $r\approx 0.74$~\AA).  Contour spacing is $\approx 0.2$~eV.}
\label{fig4}
\end{figure}

Reorienting the H$_2$ in the configuration depicted in Figs.~\ref{fig3} and \ref{fig4} (Upper Right Panels), such that the H-H bond is now perpendicular to each of the graphite sheets, did little to improve the situation.  For the configurations depicted in Fig.~\ref{fig3}, the activation barrier still remains high at ca.\ 0.5 eV (Fig.~\ref{fig3}, Lower Right Panel).  And, although we observe a significant lowering of the activation barrier to a more or less manageable ca.\ 0.3 eV in the configuration depicted in Fig.~\ref{fig4} (Lower Right Panel), there still remains the problem of how to desorb the H atoms.  The adsorption energy still remains high at ca.\ 4 to 4.5 eV (Figs.~\ref{fig3} and \ref{fig4}, Lower Right Panels).

\begin{figure}[t]
\begin{center}
\includegraphics[width=\linewidth]{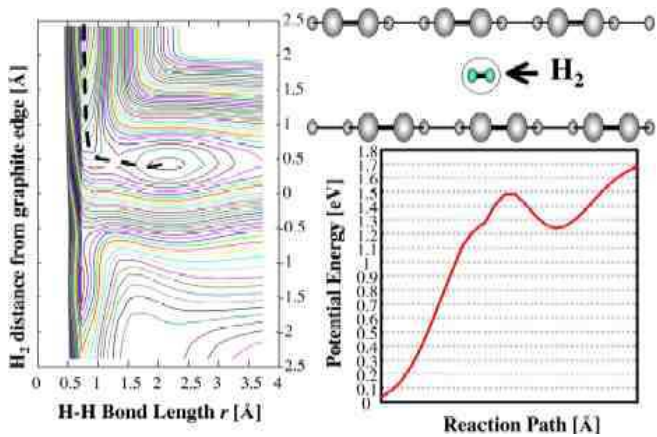}
\end{center}
\caption{({\bf Upper Right Panel}) H$_2$ configuration with respect to the armchair edge, the H-H bond is oriented parallel to and in between the two graphite sheets as indicated.  H$_2$ is constrained to dissociate over the indicated site in a parallel orientation and a planar geometry.  ({\bf Left Panel}) Contour plot of the calculated potential energy surface (PES) for H$_2$ dissociation at the indicated site on the armchair edge of graphite, as a function of the H-H separation $r$, and the normal distance $Z$ of the H$_2$ center-of-mass from the armchair edge. The dashed-line traces the {\it path of least resistance/potential} (reaction path).  The interlayer distance between graphite sheets is ca.\ 3.5~\AA.  ({\bf Lower Right Panel}) Calculated potential energy along the reaction path, i.e., along dashed-line indicated on the PES shown on the {\bf Left Panel}.  Energies are given in electronvolts relative to the values at ($Z=\infty$, $r\approx 0.74$~\AA).  Contour spacing is $\approx 0.2$~eV.}
\label{fig5}
\end{figure}

We also checked to see whether or not it would be possible to insert the H$_2$ into/between the graphite layers, as depicted in Fig.~\ref{fig5} (Upper Right Panel), and then dissociate them.   Consistent with our earlier results~\cite{h-cc}, we find that it would be difficult to dissociate H$_2$ in between graphite layers, let alone find them as H$_2$ between the graphite layers. 

These results may be compared to the activated dissociative adsorption on a graphite sheet~\cite{h2-c,h-c,h-cc} and non-activated dissociative adsorption on the zigzag edge~\cite{h-ze}.  The spectacular change in reactivity mainly lies in the difference in structure of the C atoms at these surfaces.  On a graphite sheet, all of the C atoms are bonded to three neighboring C atoms, with $sp^2$-structure.  The robustness of the $sp^2$-structure prevents the C atoms on the graphite sheet from forming new bonds.  Recall that a C atom must first form an $sp^3$-like structure before it can accomodate a H atom, directly impinging on the graphite sheet, or one coming as a product of H$_2$ dissociation~\cite{h2-c,h-c,h-cc}.  Compared to the graphite sheet, the C atoms at the outermost edges (armchair and zigzag) are bonded to only two neighboring C atoms.  This makes the edges more reactive (more susceptible to forming new bonds) to H adsorption and/or H$_2$ dissociation, as compared to the graphite sheet.  The difference in reactivity between the armchair and the zigzag edge lies in the way the C atoms at the edges are bonded.  As mentioned earlier, each of the outermost C atoms at the armchair edge has two neighboring C atoms, one of which is also located at the armchair edge.  In effect, every reaction occurring at an armchair edge, e.g., H adsorption or H$_2$ dissociation, involves an armchair edge C-C pair.  When one of the C atoms in the armchair edge C-C pair forms a new bond with the incoming H atom, the other remaining C atom (of the C-C pair) must somehow compensate for the resulting change in covalency.  Although a C atom located at the outermost part of a zigzag edge is also bonded to two neighboring C atoms,  neither of these neighboring C atoms are located at the edge.  The C atoms at the zigzag edge are, thus, already in a favorable configuration, and ready to welcome the incoming hydrogen.   (Here, we restrict the discussion to geometric effects and defer the discussion on the corresponding electronic effects~\cite{h-next}.)

In summary, based on the density functional theory, we investigate and discuss how hydrogen behaves at the edges of a graphite sheet,  in particular the armchair edge.  We found that, regardless of site, there is always an activation barrier hindering H$_2$ dissociative adsorption---a spectacular difference compared to the very reactive zigzag edge~\cite{h-ze}.  Furthermore, we also found that it was difficult to insert the whole H$_2$ through the armchair edge, in agreement with our earlier findings in our studies of H$_2$ inside/between graphite layers~\cite{h-cc}.    Based on the results we have so far gathered, we conclude that to be able to use carbon nanomaterials as a means to store hydrogen, the crucial steps would be to dissociate hydrogen first, and then (somehow) induce them to stick to the carbons on each sheets.  The results we present suggest the possible utility of the zigzag edge over the armchair edge as a reaction channel to realize efficient efficient hydrogen-storing carbon nanomaterials.
 
This work is partly supported by: a Grant-in-Aid for Scientific Research from the Ministry of Education, Culture, Sports, Science and Technology of Japan (MEXT); the MEXT Special Coordination Funds for Promoting Science and Technology (Nanospintronics Design and Realization); the 21st Century Center of Excellence (COE) Program "Core Research and Advance Education Center for Materials Science and Nano-Engineering" supported by the Japan Society for the Promotion of Science (JSPS); the New Energy and Industrial Technology Development Organization's  (NEDO) Materials and Nanotechnology program;  the Japan Science and Technology Corporation (JST) Research and Development Applying Advanced Computational Science and Technology program; and the Toyota Motor Corporation through their Cooperative Research in Advanced Science and Technology Program.  Some of the calculations were done using the computer facilities of JST, the Yukawa Institute Computer Facility (Kyoto University), the Institute for Solid State Physics (ISSP) Supercomputer Center (University of Tokyo) and the Japan Atomic Energy Research Institute (JAERI).

\end{document}